\documentclass[preprint,12pt]{elsarticle}

\usepackage{amssymb}
\usepackage{txfonts}
\usepackage{mathbbol}
\usepackage{amsfonts}
\usepackage{mathrsfs}
\usepackage{epsfig,bm,dcolumn}
\usepackage{graphicx}
\usepackage{color}
\usepackage{overpic}
\usepackage{slashed}
\usepackage{hyperref}
\usepackage{subfig}
\usepackage[capitalise]{cleveref}

\journal{Physica A}

\begin{document}

\begin{frontmatter}



\title{Escape dynamics based on bounded rationality}

\author[wang]{Lingxiao Wang\corref{cor1}}
\ead{wlx15@mails.tsinghua.edu.cn}
\ead[url]{https://sites.google.com/view/lingxiao}
\cortext[cor1]{Corresponding author. Tel.: +86  010-62772694.}
\author[jiang]{Yin Jiang}
\address[wang]{Department of Physics, Tsinghua University and Collaborative Innovation Center of Quantum Matter, Beijing, 100084, PR China}
\address[jiang]{Department of Physics, Beihang University, Beijing, 100191, PR China}

\begin{abstract}
The bounded rationality plays a vital role in the collective behavior of the evacuation process. Also investigating human behavior in such an extreme situation is a continuing concern within social psychology. In this paper, we construct a cellular automaton (CA) model for the escape dynamics, and the bounded rational behavior induced by heterogeneous information is introduced. The non-trivial behavior shows in the replicator dynamics method with mean field approximation, where people's perception of the distribution of population and velocity is reduced to an average value in a certain direction. Analyzing the escape efficiency shows that under the premise of rationality, the bounded rational strategy can get higher performance. Interestingly, a quantifiable  meta-stable state appears in the escape process, and the escape time is power-law dependent on system size.

\end{abstract}


\end{frontmatter}

\section{Introduction}
Public security is the cornerstone of national and social stability.  In addition to the direct loss of life and property caused by natural disasters, crowd congestion in emergencies often leads to disaster (e.g. clogging and stampede\cite{helbing:2000simulating,hughes:2002continuum,helbing:2005selforganized}) . It naturally becomes important to understand the collective behavior patterns in case of emergency. Recent models and experiments show that the characteristics of group movement emerge as short intermittent bursts\cite{pastor:2015experimental,nicolas:2018counterintuitive,bain:2019dynamic}. When the desire of escaping danger exceeds the idea of avoiding the collision, people's behavior pattern changes from order to disorder.  However, our understanding of the transition is still limited. Even the research on escape dynamics in emergencies has a long history\cite{togawa:1955study,henderson:1971statistics}, it was not until the 1990s that the dynamics on collective behavior attracts people's attention\cite{helbing:2005selforganized,helbing:1995social}. Subsequently, game theory, decision theory, communication model and queuing model had also been comprehensively applied\cite{vermuyten:2016review}. However, due to the lack of individual self-organization, the prediction of many results deviates from the actual situation. Some works start from the hydrodynamic model to study the collective behavior of the population\cite{low:2000statistical,hughes:2002continuum}. It successfully explained the group behavior during the pilgrimage to Mecca. Also, we can get some non-trivial macroscopic patterns in these hydrodynamics models\cite{bain:2019dynamic,nicolas:2019mechanical}. However, the macro-model sometimes is coarse and will miss individual information, while the micro-model represented by the Social force model, the Cellular automata model, and the Magnetic field force model\cite{helbing:1995social,helbing:2000simulating,burstedde:2001simulation,weng:2006cellular,guo:2012heterogeneous} can give a more completed description for the escape dynamics. Simulating the collective behavior in emergencies becomes more convenient with the increasing of computer capabilities.  The simulating results, such as the "faster-is slower" phenomenon also can be verified by experiments on the different group (e.g people, vehicles, ants, sheep, microbial populations, etc.)\cite{pastor:2015experimental,patterson:2017clogging,aguilar:2018collective,delarue:2016selfdriven,garcimartin:2015flow}.

With the development of sensor technology and the improvement of microchip computing power, collecting more redundant data and using more realistic methods to simulate the escape process becomes possible\cite{moussaid:2011how,corbetta:2017fluctuations,helbing:2011recognition,zanlungo:2017intrinsic,wang:2018study}. It is of practical significance that the disaster may happen in different complex en-vironments\cite{shi:2018examining}. 
{It's natural to introduce the game theory into the evacuation model, for it can include the interaction between people (and environment) self-consistently\cite{kulkarni:2019sparse,dongmei:2017dynamics}. The embedded game action occurs in conflicts mainly, and people will decide to advance or avoid others by pay-off matrix\cite{hao:2011pedestrian,zheng:2011conflict,zheng:2011modeling,shi:2013evacuation,guan:2016cellular}. However, handling conflicts are only parts of escape dynamics, how human beings make decisions induced by specific circumstances also is an essential part\cite{vonschantz:2015spatial}. In this paper, we introduce the replicator dynamics into the evacuation. It makes people's response to the external environment can be included in the decision-making process.}

Besides, some researches focus on the behavior itself, for escape dynamics provides an extreme case to investigate collective behavior\cite{corbetta:2017fluctuations,guo:2012heterogeneous,nicolas:2018trap,cavagna:2018physics}. The diverse and fascinating collective behaviors occur in both virtual and real space\cite{castellano:2009statistical,ball:2012why}: social network, financial network and social norms, these virtual social connections naturally incubate the collective behavior; as for the real space, collective modes are common in urban dynamics, traffic flow, and pedestrian dynamics.
Therefore, the escape dynamics provides us with an extreme scene, in which human instinct dominates\cite{nowak:2005emergence,zanlungo:2017intrinsic}. It makes us have chances to effectively study human behavior itself without complex social relations.

Based on the point, this work introduces bounded rationality\cite{pan:2014spatial,simon:1983reason}from behavioral economics in the escape dynamics through the replicator dynamics method\cite{heliovaara:2013patient,taylor:1978evolutionary}.  A cellular automaton model is used to model the escape dynamics in a closed boundary. And the influence of boundedly rational behavior strategy on collective behavior is investigated by using mean field approximation\cite{lasry:2007mean}. Escape efficiency is affected by the environment and heterogeneous information processing. We also analyze the group and individual escape time, giving a possible picture to understand the connection between collective behavior and individual action.

\section{Methods}
\subsection{Cellular Automaton Model}
We construct a cellular automation model for simulating the pedestrian flow in a two-dimensional system. The underlying structure is a $ L\times L $ cell grid, where $ L $ is the system size. The state of cell can be empty, or occupied by one pedestrian exactly or wall. It's an instructive sample, once we set the size of cell as  0.5m $\times$ 0.5m, which can simulate the escaping when some disaster happens. Model adopts the Moore neighbor, and pedestrians update their positions by transition matrices $ T(i,t) $,
\begin{equation}
	T(i,t)=\left( \begin{array}{ccc}
		P_{1,1}(i,t) & P_{1,2}(i,t) & P_{1,3}(i,t) \\ 
		P_{2,1}(i,t) & P_{2,2}(i,t) & P_{2,3}(i,t) \\ 
		P_{3,1}(i,t) & P_{3,2}(i,t) & P_{3,3}(i,t)
	\end{array} \right) 
\end{equation} 
where $ P_{m,n}(i,t) $ means the possibility that the pedestrian $ i $ moves from $ t $ time at position $ (x(i,t),y(i,t) ) $ to next time-step position. The  neighbors' directions are labeled by $ (m,n)$, where $ m,n=1,2,3 $. Each cell can either be empty, or occupied by wall or exactly one pedestrian. Every time-step pedestrian can choose to move into a new position or stop. Once we have chosen the location of the exit, the synchronously updated cellular automaton can imitate the escape process\cite{burstedde:2001simulation,kirchner:2003friction}.
\subsection{Heterogeneous Information and Bounded Rationality}
Bounded rationality is formalized with such major variables as incomplete information, information processing and the non-traditional policy-maker target function\cite{simon:1983reason}. Heterogeneous information could be the reason why people shows irrationality\cite{lee:2001effects,nowak:2005emergence,moussaid:2011how,heliovaara:2013patient}. The extreme situation of escaping from disasters constrains people's behavior, in which only intuition or social habits remains, no long term trade-off. The replicator dynamics modeling\cite{heliovaara:2013patient,taylor:1978evolutionary} can link the different behaviors, whether practical or spiritual, during the escaping process. The  transition possibility $ P(i,t) $ derives from the follow definition,
\begin{equation}
	P_{m,n}(i,t)=\frac{B_{m,n}(i,t)R_{m,n}(i,t)}{\sum B(i,t)R(i,t)}
\end{equation}
where $ R(i,t), B(i,t) $ means the weight from rational and bounded rational part respectively. The definition of the components in matrix $R_{m,n}(i,t)=O_{m,n}(i,t)E_{m,n}(i,t) $,
\begin{equation}
	O_{m,n}(i,t)=\left\{
	\begin{array}{rcl}
		1 &  & {empty} \\ 
		\epsilon &  & {occupied}
	\end{array},  E_{m,n}(i,t)=\Bigg\{\begin{array}{rcl}
		\alpha &  & {exit} \\ 
		\epsilon &  & {nothing}
	\end{array}\right.
\end{equation}
which means if the position $ (m,n) $ around the individual $ i $ at $ t $ time is empty, the $ O_{m,n}(i,t)=1 $, whereas the value is $ \epsilon $. And the $ E_{m,n}(i,t)=\alpha $ only holds when the exit direction is indicated by $ (m,n) $, if not take the value $ \epsilon $. The $ \epsilon $ is a minimum value that the calculation accuracy can reach. The parameter $ \alpha $ represents  the attraction of the exit to persons want to escape, or the importance of the information of the exit position.

The definition of the bounded rational part $ B_{m, n}$ relies on the heterogeneous information from the crowd. As the transport model of statistical physics inspired us, the escape dynamics needs more information that persons' position and velocity distribution, the basic variables of the transport theory. Considering the full information cannot easily be achieved by individuals, the mean-field approximate can provide a global perception for the people on move, which shows as the follow,
\begin{equation}
	B_{m,n}(i,t)=\left\{
	\begin{array}{rcl}
		1 &  & {rational} \\ 
		n_{m,n}(i,t)&  & {crowd}\\ 
		v_{m,n}(i,t)&  & {follower}
	\end{array}\right.
\end{equation}

The $ rational $ indicates the transition possibility only decided by $ R(i,t) $, the neighbor occupied state and the direction of exit, or the objective environment. The $ crowd $ defines $ n_{m,n}(i,t)={\sum_{m,n}N(i, t)}/{\sum_{All} N(i, t)} $, where $ N(i, t) $ is the population distribution at $ t $ time. The definition shows the proportion of individuals in $ (m,n) $ orientation as mean-field approximation, and people will be attracted to the direction with more density. We use it to mimic the "crowd" behavior for individuals, which also means people can potentially get more population density information. As for the $ follower $, $ v_{m,n}(i,t)=\sum_{j} N_{\vec{v}(j, t)=(m, n)}/{\sum_{All} N(i, t)}$, where $ \vec{v}(j, t) $ is the velocity distribution at $ t $ time. The proportion of individuals move to $ (m,n) $ orientation has been extracted, and people will follow others as the weight. It transfers more potential velocity information to people. 
{The latter two strategies show the individual can process the heterogeneous information -- population and velocity of all persons. The population $n(x,y,t)$ and velocity $v(x,y,t)$ of the crowd are the global continuous distribution quantity in reality, which affects the human behavior indirectly, for people can gather and process information from the environment\cite{moussaid:2011how,lee:2001effects}. In this work, the distribution is discrete and the individual can process them as background, that's what above definition means. People's perception to the distribution is reduced to the average value in a certain direction, a mean background field, as what  statistical physics did in a many-body system.}

\subsection{Evolution Rules}
The model escape rules gives as follows,

\textbf{Step.1 Initialization.} Set the position of exit $ (x, y) $ and generate $ N(i,t) $ population distribution at the $ L\times L $ lattice. At the $ t=0 $ time, disaster turns out and individuals begin to move;

\textbf{Step.2 Evolution.} At the $ t $ time step, the individual $ i $ move to the next position as transition matrices $ T(i,t) $ at $ t+1 $ time step. Update all individuals synchronously, and the conflict will be handled by compared the transition possibility;

{\emph{Handle Conflicts}. The conflicts occurs when the two or more persons want to move into the same position, and what we do to handle the conflicts is to compare their transition possibilities $ P_{m,n}(i,t)$ which reflects their willingness to move. For example, the individual $j$ and $k$ both want to move into position $(x,y)$, and the corresponding possibility for the $j$ is  $ P_{m,n}(j,t)$ and the $k$ is $ P_{m',n'}(k,t)$. If the $ P_{m,n}(i,t)>P_{m',n'}(k,t)$, then the individual $j$ move successfully and the $k$ stayed where it was, and vice versa. For equal cases, one is randomly selected. It can be easily extended to the situation of many people. }

\textbf{Step.3 Escape.} For the individual whose destination is exit at the next time step, escape successfully, and remove it from the lattice and reduce population as $ N(t+1)=N(t)-1 $. If $ N(t)=0 $ the escape stops.

\textbf{Step.4 Update.} Update the transition matrices as above strategies, turn to \textbf{Step.2}. The $ t $ time step escape finished.

\section{Results and Discussions}
\subsection{Dynamics Simulation}
\begin{figure}[h]
	\centering
	\includegraphics[width=4.8cm]{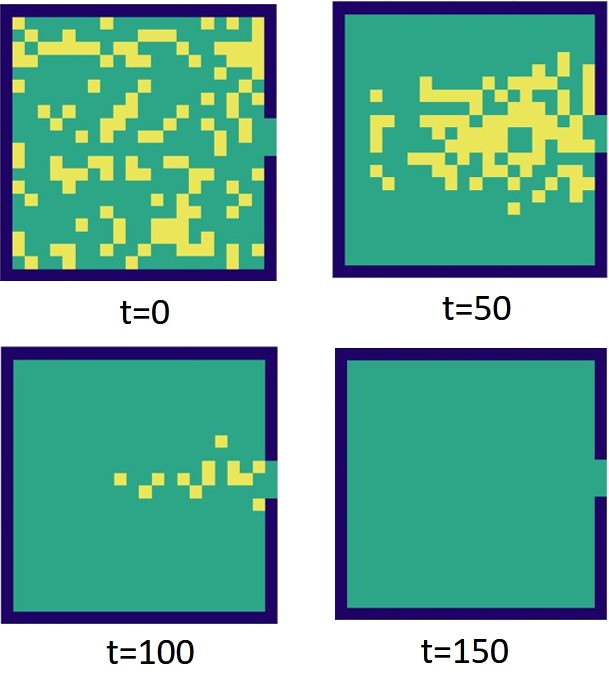}
	\caption{\textbf{Evolution Sample.} Initial population ratio $ \rho_0 \approx0.3$, lattice size $ L=20 $, the exit size is 2, and the parameter $ \alpha=10 $ with the $ rational $ strategy.}
	\label{sample}
\end{figure}
Firstly we build an escape dynamics simulation frame based on the escape rules, in which three different strategies for processing heterogeneous information have been added. As an example, \cref{sample} depicts a typical escape process: at the beginning individuals distribute in the lattice $ L\times L $ randomly with initial population ratio $\rho_0 $; then they move into the exit direction as the parameter $ \alpha $ which shows how important the exit information is for them; at the $ t=150 $ time step, people escape from the disaster area.

\begin{figure}[h]
	\centering
	\includegraphics[width=6cm]{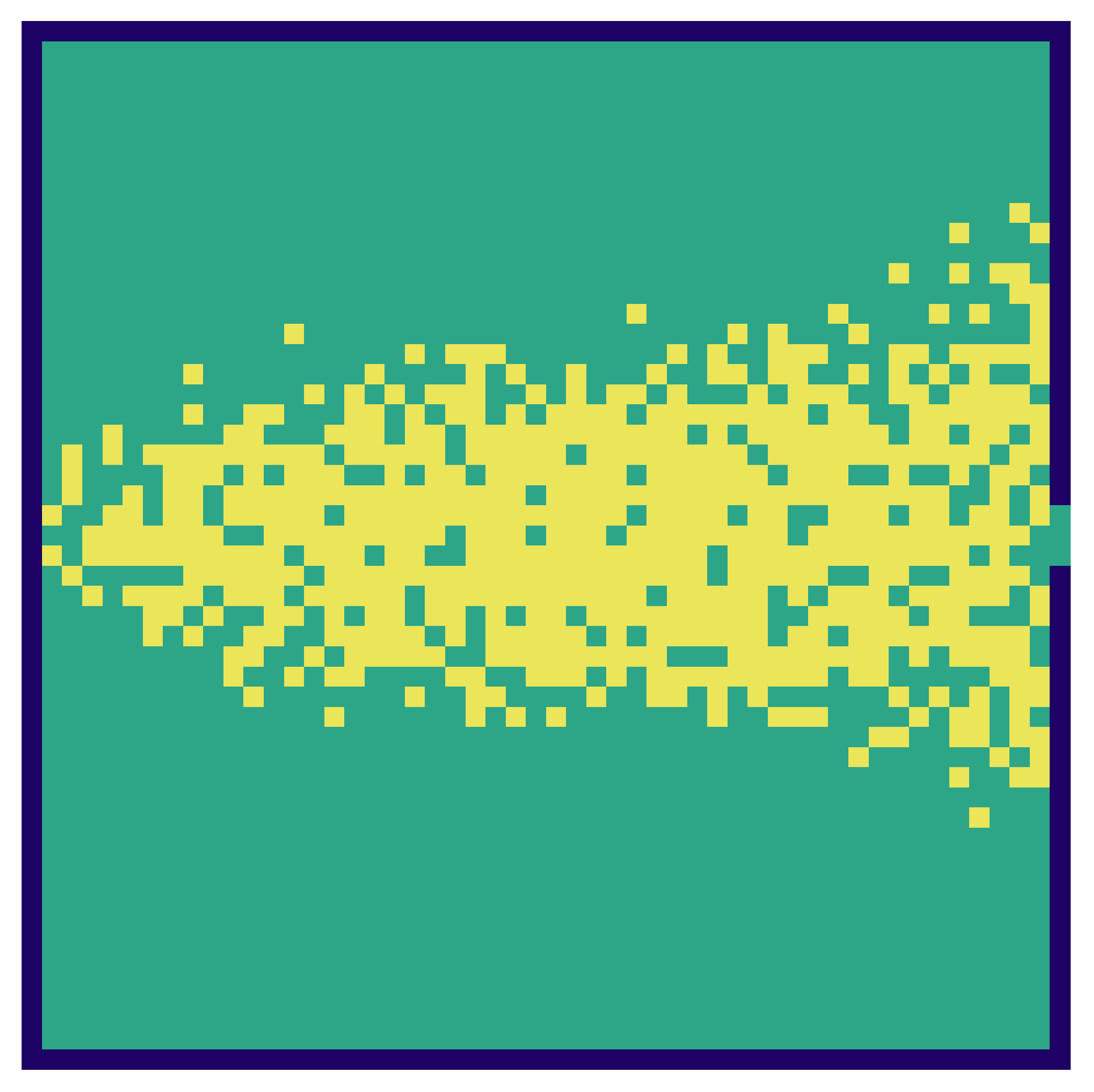}
	\caption{\textbf{Arch-like blocking.} Initial population ratio $ \rho_0 \approx0.3$, lattice size $ L=50 $,  the exit size is 2, and the parameter $ \alpha=10 $,  at time step $ t=80 $ with the $ rational $ strategy.}
	\label{arch}
\end{figure}
In the \cref{arch}, we show the arch-like blocking as the other simulations and experiments found\cite{pastor:2015experimental,patterson:2017clogging,delarue:2016selfdriven,garcimartin:2015flow}, which indicates the escape dynamical model catches the key points for the flock clogging problem.

\begin{figure}[h]
	\centering
	\subfloat[$\rho_0\approx0.12$, $ \alpha=10$]{\includegraphics[width=6cm]{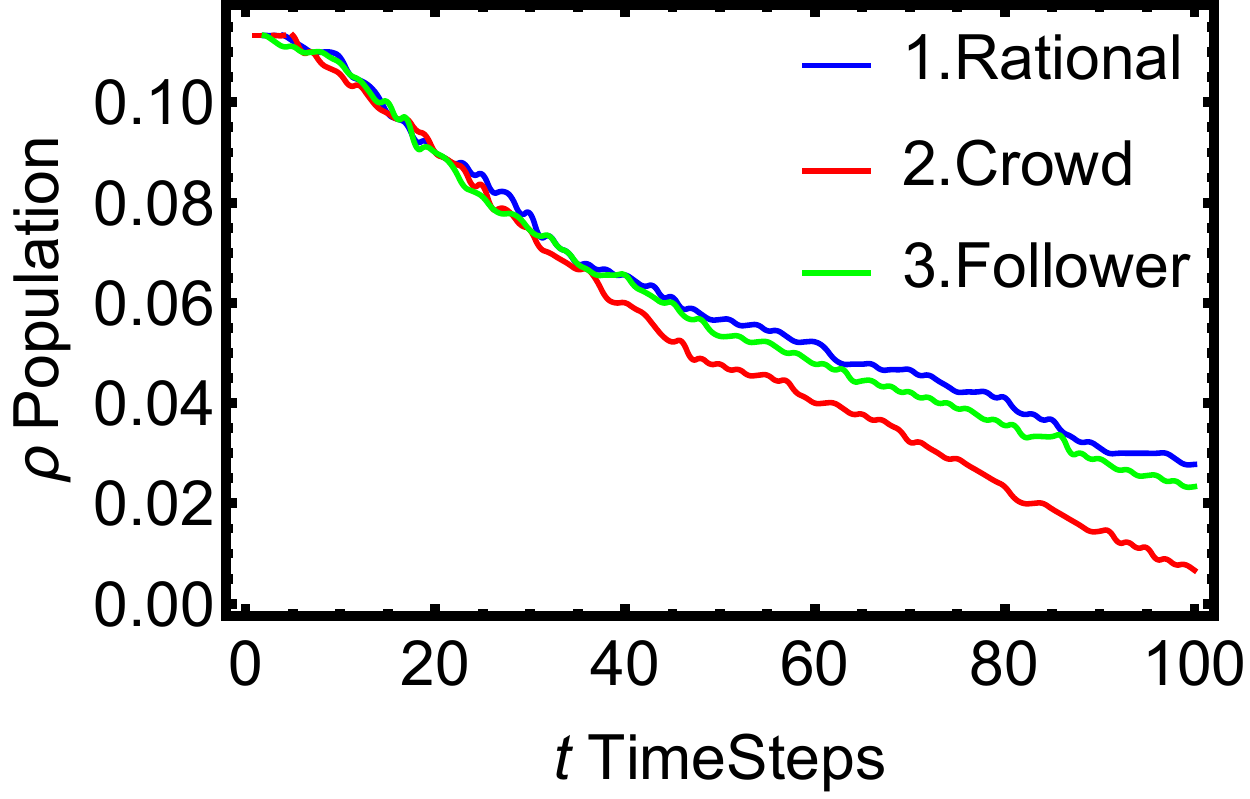}
		\label{rho1}}
	\hspace{0.1cm} 
	\centering
	\subfloat[$\rho_0\approx0.51$, $ \alpha=10$]{\includegraphics[width=6cm]{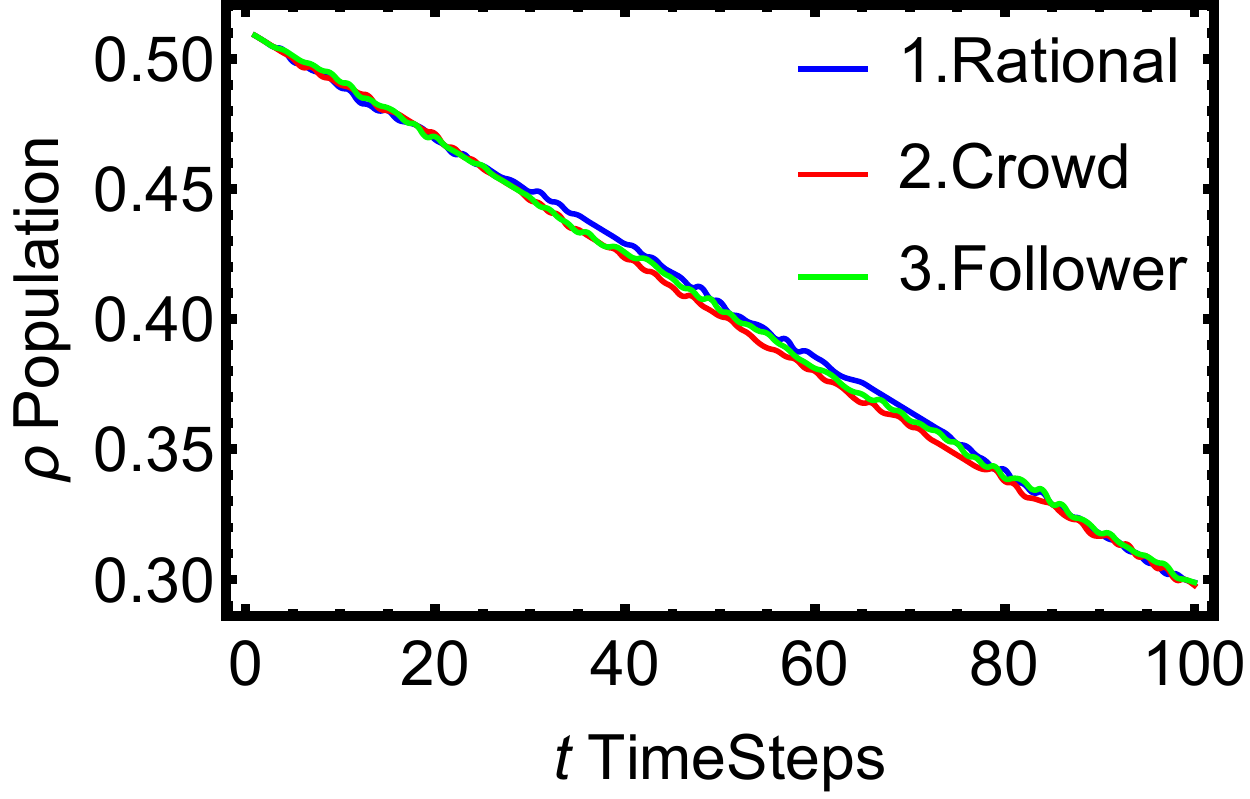}		\label{rho5}}
	\\
	\centering
	\subfloat[$\rho_0\approx0.3$, $ \alpha=10$]{\includegraphics[width=6cm]{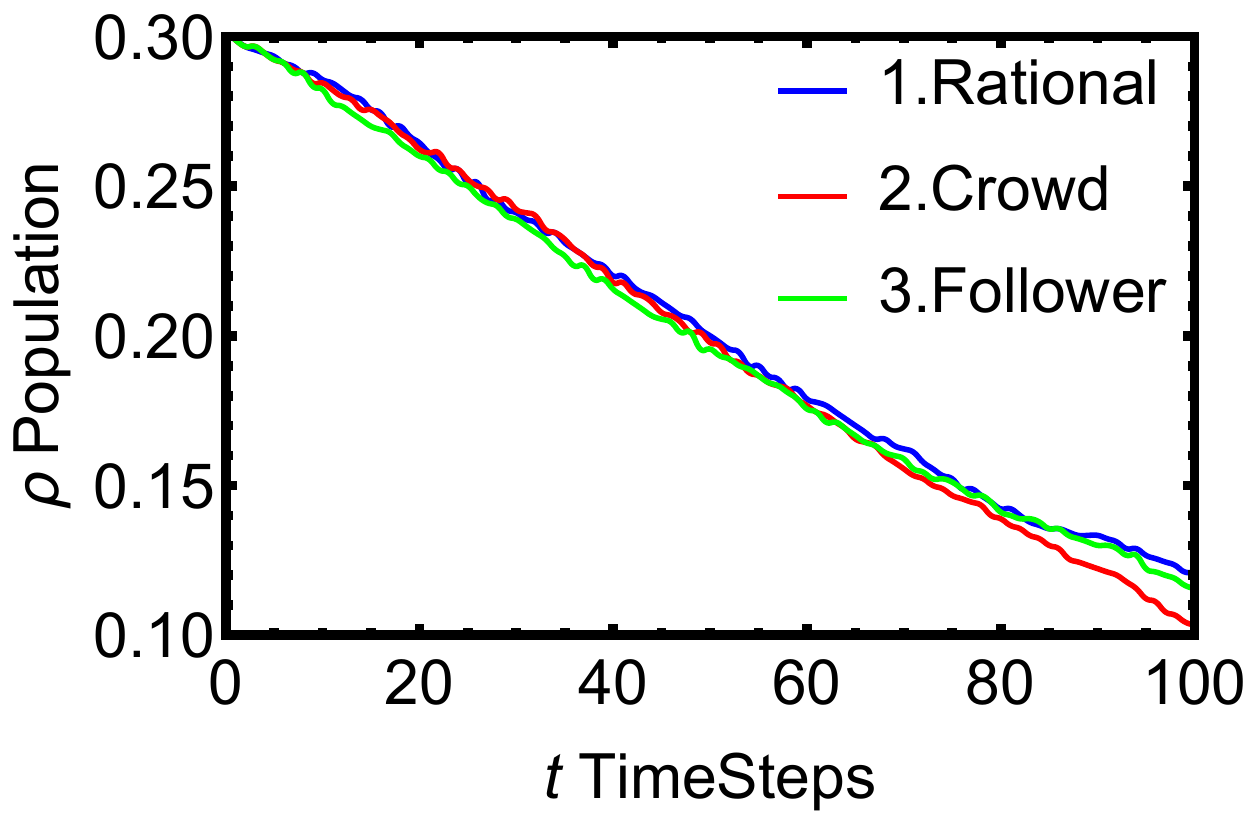}		\label{alpha10}}
	\hspace{0.1cm} 
	\centering
	\subfloat[$\rho_0\approx0.3$, $ \alpha=1$]{\includegraphics[width=6cm]{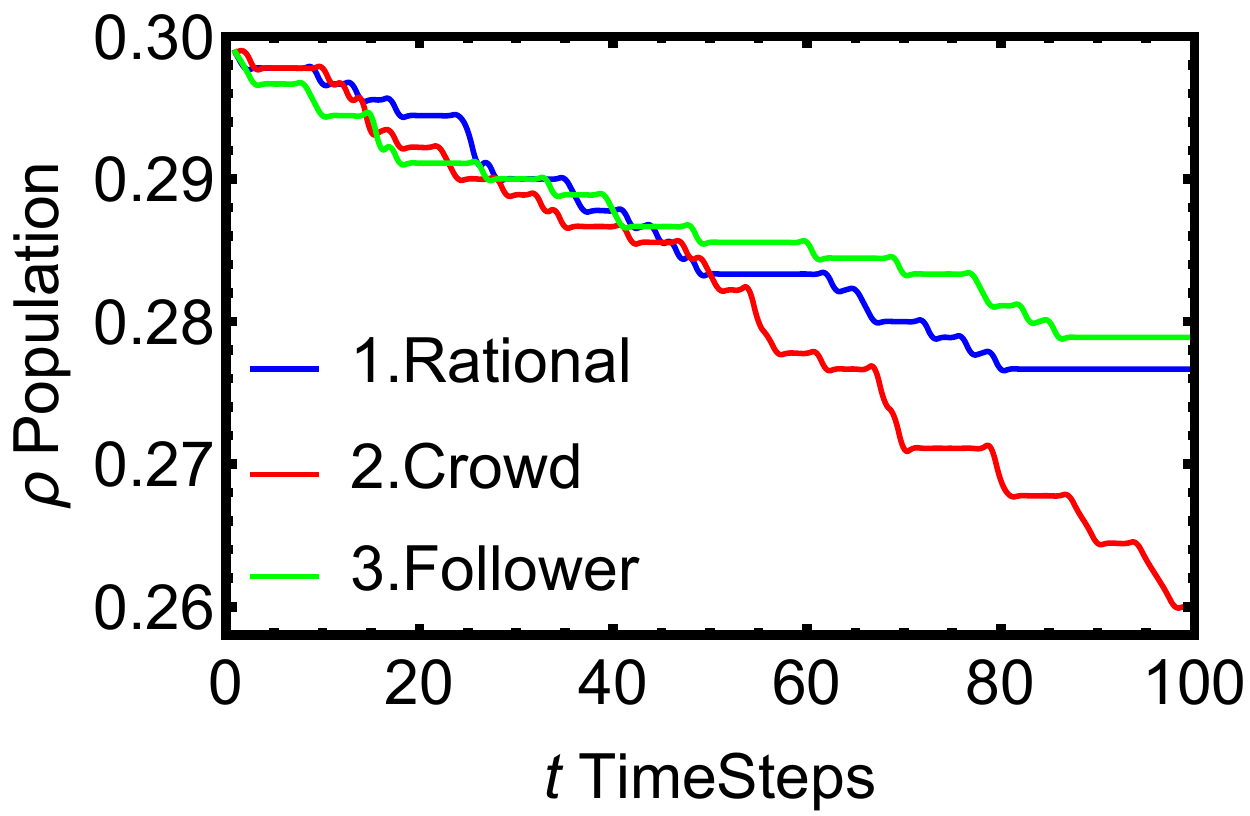}		\label{alpha1}}
	\caption{Escape Efficiency}
	\label{efficiency}
\end{figure}
\cref{efficiency} Shows the escape efficiency in the $ t\in[0,100]$ time steps at the parameter, lattice size $ L=30 $ and the exit size is 2. Over time, all three behavior strategies are driving groups to flee disaster areas.~\cref{rho1,rho5,alpha10} have different $ \rho_0 $ and the high initial population case can escape faster than the lower. It's because the low initial population case has more empty space, which reduces the escape chances. As for the three behavior strategies, they have similar escape efficiency in the higher $ \rho_0 $ case, where the population decreasing is almost same in 100 time steps. The three behavior strategies show relative large differences in low density situations. During the same time, the $ crowd $ made more people escape, the $ follower $ followed, and the relative worst is the $ rational $ strategy. The above results illustrate that the "Bounded Rationality" strategies can refine information from environment, which makes people work better in low density case.~\cref{alpha10,alpha1} have different $ \alpha $ and the high $ \alpha $ case can escape faster than the lower. It makes sense that $ \alpha $ represents the importance of exit, and it could be the intensity of exit signs or how clear people knows the exit information. In the \cref{alpha1}, the exit information parameter $ \alpha=1 $, it makes people random walks at some time and the $ crowd $ strategy shows more powerful capability of driving people to the exit.

\subsection{Collective Behavior and Meta-stable State}
\begin{figure}[h]
	\centering
	\includegraphics[width=8cm]{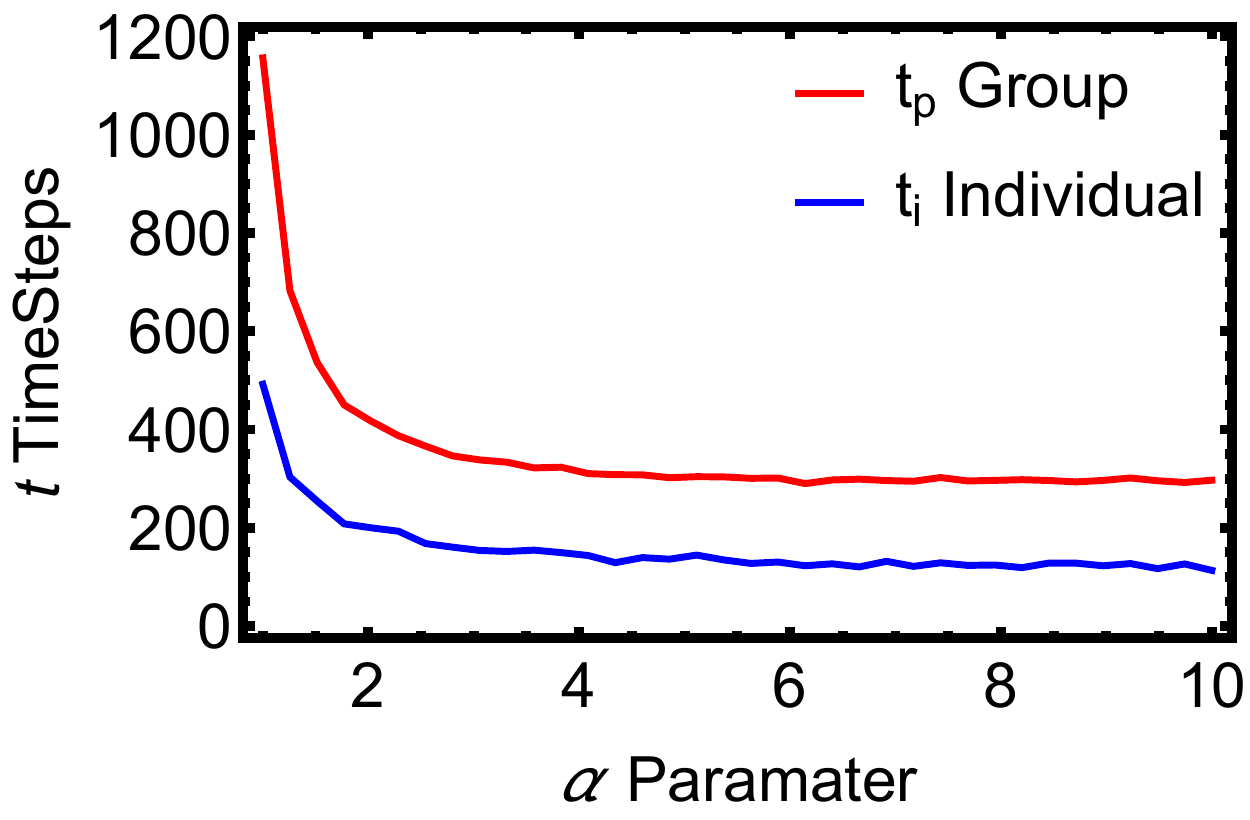}
	\caption{The escape time is decreasing with $ \alpha $ at $ p=0.9 $ }
	\label{alpha}
\end{figure}
\begin{figure}[h]
	\centering
	\includegraphics[width=8cm]{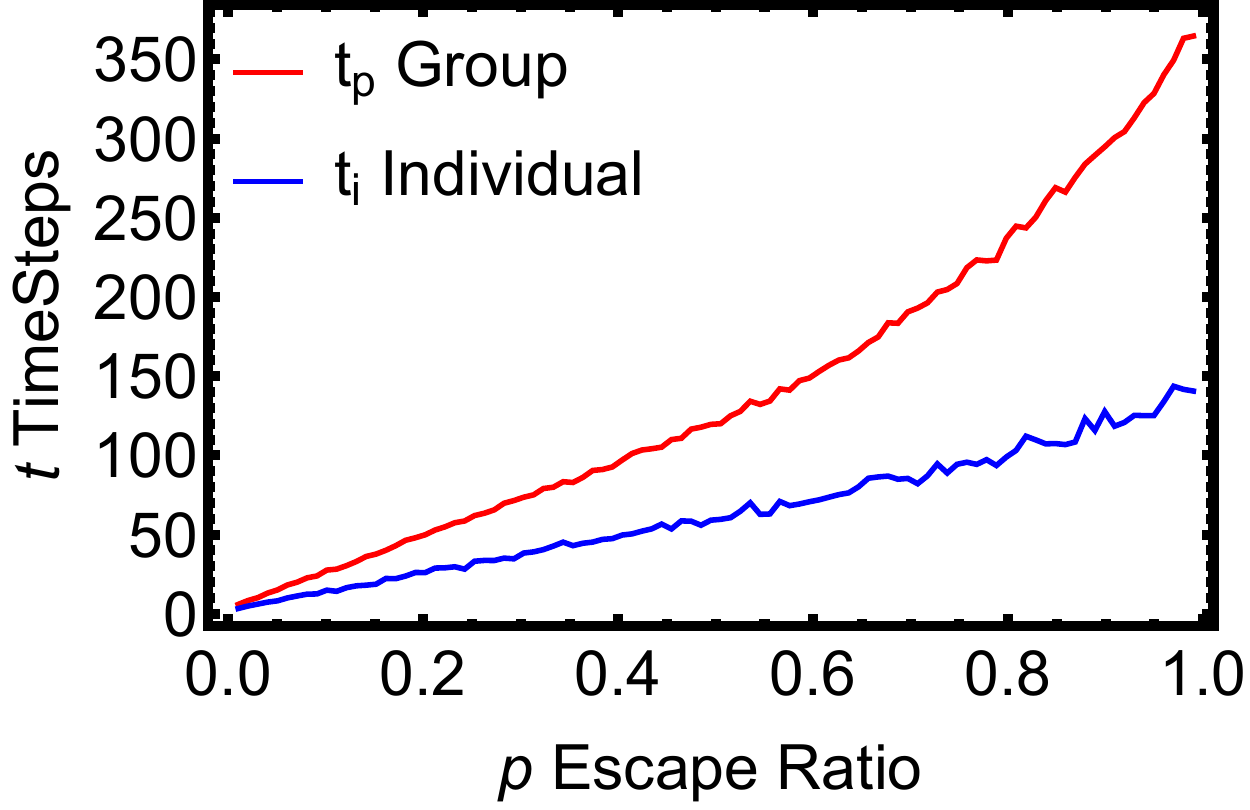}
	\caption{The escape time is increasing with $ p $ at $ \alpha=10 $}
	\label{escaperatio}
\end{figure}
To investigate the factors affecting collective behavior more quantitatively, we define the escape ratio $ p $, the corresponding group escape time $ t_p $  and individual escape time $ t_i $. The $ t_p $ is the time taken to evacuate $ p $ of the population and the $ t_i=\frac{\sum^{N(t_p)}_{n=1}t_n}{N(0)\times p}$ is the average time, where the $ t_n $ is the time taken to evacuate $ n $th person. \cref{alpha,escaperatio} shows the escape time in $ crowd $ strategy at different parameter $ (\alpha, p) $, lattice size $ L=30 $, initial population ratio $ \rho_0=0.5 $ and the exit size is 2. The parameter $ \alpha $ has played a key role in the escape dynamics, as~\cref{alpha} shows, the escape time decays exponentially with this exit parameter. At small $ \alpha $ region, the $ \alpha $ increasing would push people escape faster; at the large region, the effect of increasing is limited. The results reflect that there is a effective range for the relative importance of exit information in the dynamics processing. The other statistical result gets in \cref{escaperatio}, where group escape time $ t_p $ increases with escape ratio near exponentially and the individual escape time $ t_i $ increases near linearly. It implies that the time taken to rescue more people rise sharply, which reveals the conflict of interest between individual and group. The above results are independent of initial conditions, for every case runs 20 times in different random initial population distribution.

\begin{figure}
	\centering
	\includegraphics[width=10cm]{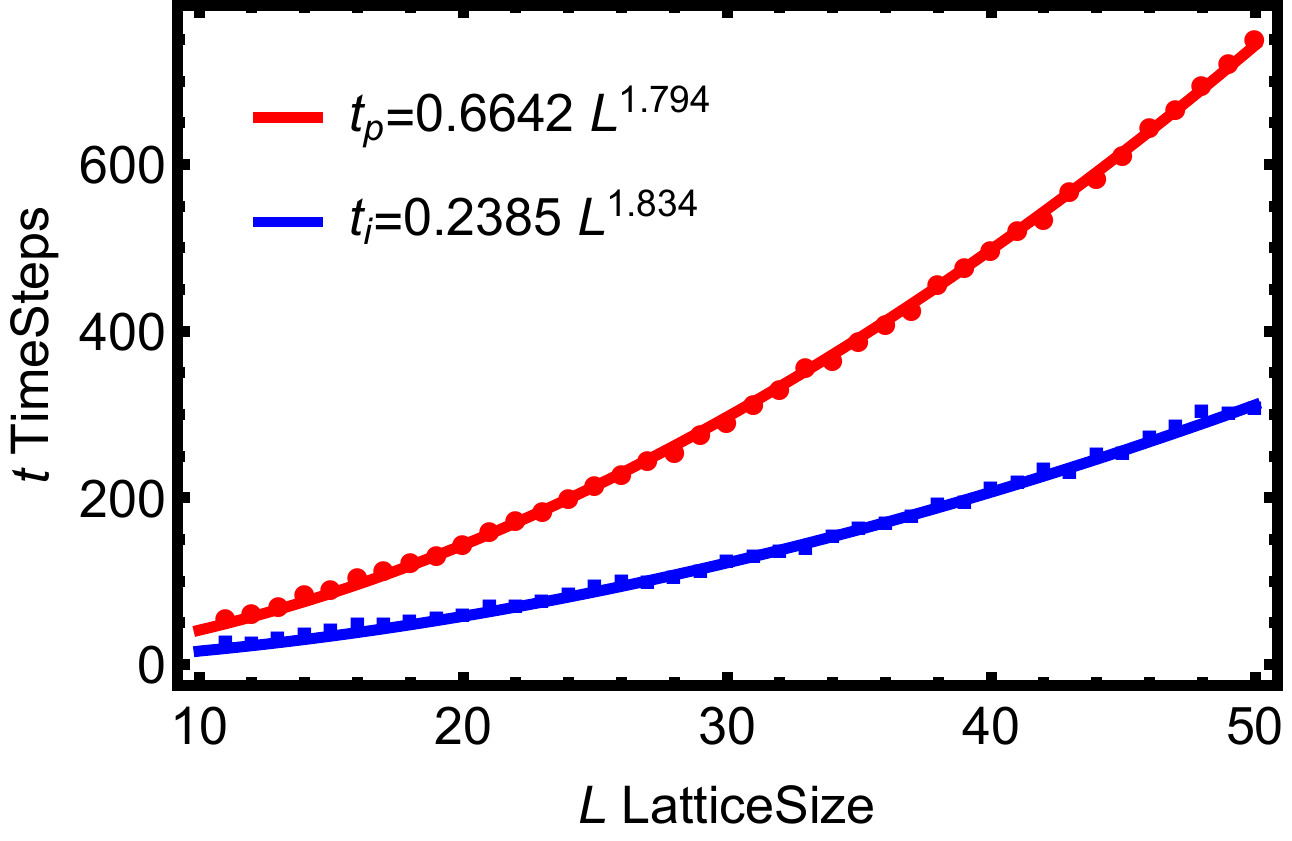}
	\caption{The scale dependent of escape time}
	\label{size}
\end{figure}
\begin{figure}[htbp]
	\centering
	\includegraphics[width=10cm]{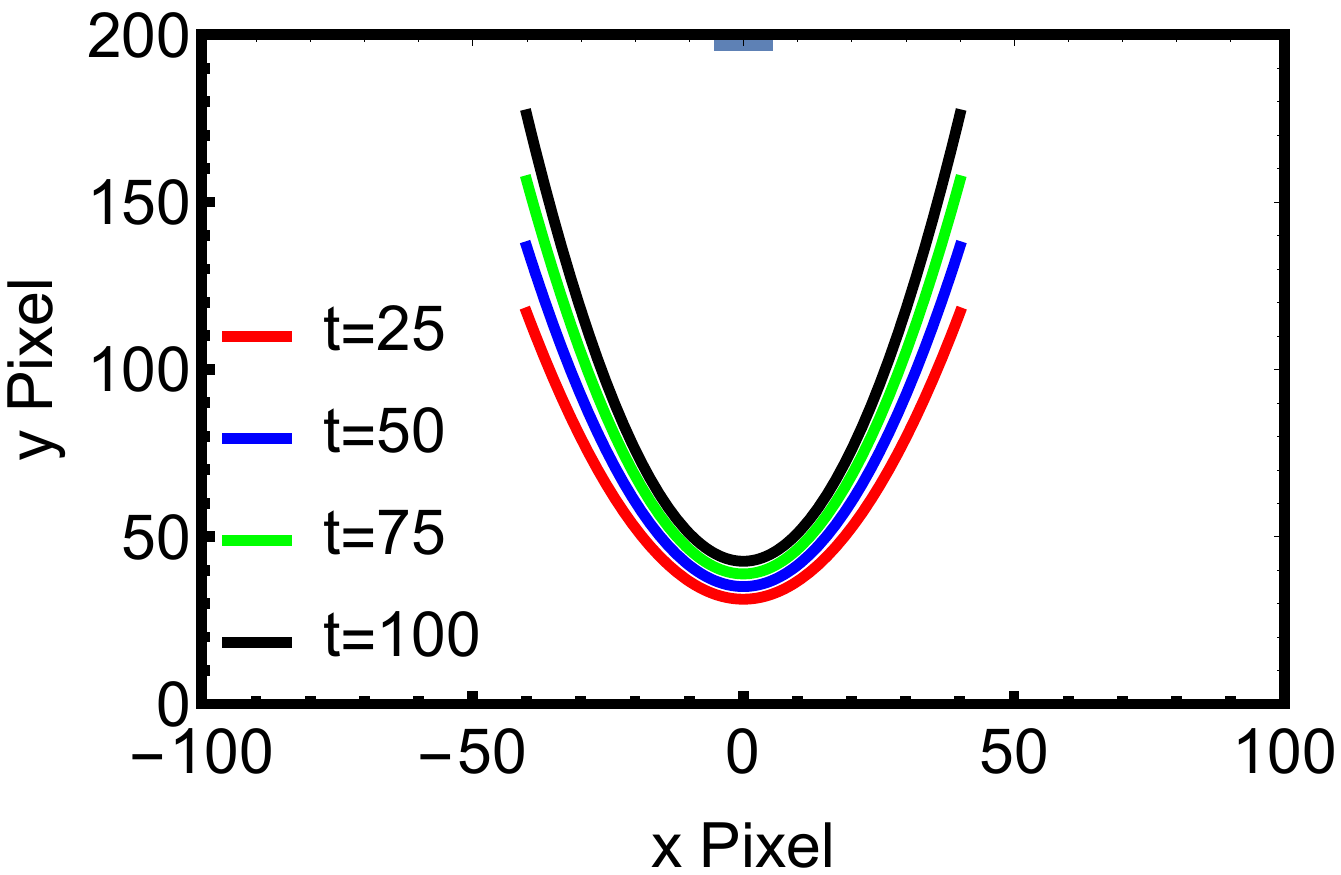}
	\caption{The parabola fitting for the edge}
	\label{parabola}
\end{figure}
The above nonlinear dependence inspired us to study the influence of system size on escape time. The results exhibit in \cref{size}, the other system parameters are: initial population ratio $ \rho_0=0.5 $, the parameter $ \alpha=10, p=0.9 $ and the exit size is same as before. Both the group and individual escape time have $ \ln(t_{p/i})/\ln(L)\approx 1.8 $ , and the exponential coefficient deviating from system area $ L^2 $ slightly. It can be treated as a signal of the critical self-organizing behavior for the escape dynamics\cite{castellano:2009statistical,cavagna:2018physics}. At the same time, we notice that the arch-like blocking is a corresponding phenomenon during the critical process. It can be understood as a meta-stable state, for the empty is the stable final state. We extract the edge curve as parabola fitting ($y= a_2(t)x^2+a_1(t) $) from the escape maps (e.g.\cref{parabola} as a sample at $ L=40, \alpha=10, \rho_0=0.3$ ), and this meta-stable state will emerge in the mid-time term ($ t\in[20,120]$ for this sample). Where the pixel positions of corresponding evolution patterns represented by coordinate axes (x, y), and the position of exit is (0,200). As \cref{parabola} shows, the edge of population will push forward and become narrower. It can be described by the two fitting parameters: $ a_1(t)=0.1512 t+27.53$, $a_2(t)=3.986\times10^{-4} t+0.04354$ in this sample. As time goes, the opening of parabola becomes narrow until the meta-stable state disappears. Even parabola fitting is not enough accurate, it still reflects such a meta-stable propulsion process.

\section{Conclusions}
In this work, we construct an escape dynamics frame with cellular automata model and can generate the space-time escape map similar to the actual situation. We also use the Replicator Dynamics to combine the bounded rational behavior into the escape. And three different behavior strategies were compared: the $ rational $, $ crowd $ and $ follower $. The difference among these strategies lies in the completeness of the information.  Our results show that under the premise of rationality, the bounded rational behavior of the $ crowd $ can get higher evacuation efficiency. Subsequently, the influence of escape ratio, exit parameter and system size on the collective behavior pattern was further studied by using a parametric method in the $ crowd $ behavior strategy. We also introduced group escape time $ t_p $ and average escape time $ t_i $ to study the behavior patterns of people fleeing disaster areas and found that changes in the external environment and individual rationality will have non-trivial effects on it. The increase in the importance of exit information $ \alpha $ will help to improve the efficiency of escape, while conflicts of interests between individuals and group occur in the process of increasing the escape ratio $ p $. Moreover, escape time is power-law dependent on system size. It can be treated as a signal of the critical self-organizing behavior for the escape dynamics. Finally, we extract the edge curve as parabola fitting from the escape dynamics, which reflects such a meta-stable propulsion process.

In the extreme case, this work uses the replicator dynamics for reference, introduces bounded rationality, and adopts mean-field approximation to study collective behavior. Furthermore, we can use more abundant tools to carry out more in-depth research on such problems: for example, using the state-of-the-art deep learning methods to recognize potential collective behavior and avoid trampling;  make online games~\cite{mao:2016experimental,awad:2018moral} based on the model, measuring the parameter $ \alpha $ as a group rational scale;  search the possible order parameter during such a emergence phenomena, etc. Overall, this work has provided a deeper insight into the correlation between human rationality and corresponding collective behaviors.

\section{Acknowledgments}
The authors thank Prof.WengGuo Weng and Dr.Xingyu Guo for useful discussions and comments. The work is supported by the Qingfeng Scholarship from Tsinghua University. LX and YJ are also supported by the China Scholarship Council(CSC) for visiting at the University of Tokyo and the Zhuobai Program of Beihang University respectively.

\bibliographystyle{elsarticle-num}
\bibliography{escapedynamics_physica}

\begin{thebibliography}{10}
\expandafter\ifx\csname url\endcsname\relax
  \def\url#1{\texttt{#1}}\fi
\expandafter\ifx\csname urlprefix\endcsname\relax\def\urlprefix{URL }\fi
\expandafter\ifx\csname href\endcsname\relax
  \def\href#1#2{#2} \def\path#1{#1}\fi

\bibitem{helbing:2000simulating}
D.~Helbing, I.~Farkas, T.~Vicsek, Simulating dynamical features of escape
  panic, Nature 407~(6803) (2000) 487--490 (Sep. 2000).
\newblock \href {https://doi.org/10.1038/35035023}
  {\path{doi:10.1038/35035023}}.

\bibitem{hughes:2002continuum}
R.~L. Hughes, A continuum theory for the flow of pedestrians, Transportation
  Research Part B: Methodological 36~(6) (2002) 507--535 (Jul. 2002).
\newblock \href {https://doi.org/10.1016/S0191-2615(01)00015-7}
  {\path{doi:10.1016/S0191-2615(01)00015-7}}.

\bibitem{helbing:2005selforganized}
D.~Helbing, L.~Buzna, A.~Johansson, T.~Werner, Self-{{Organized Pedestrian
  Crowd Dynamics}}: {{Experiments}}, {{Simulations}}, and {{Design Solutions}},
  Transportation Science 39~(1) (2005) 1--24 (Feb. 2005).
\newblock \href {https://doi.org/10.1287/trsc.1040.0108}
  {\path{doi:10.1287/trsc.1040.0108}}.

\bibitem{pastor:2015experimental}
J.~M. Pastor, A.~Garcimart\'in, P.~A. Gago, J.~P. Peralta,
  C.~{Mart\'in-G\'omez}, L.~M. Ferrer, D.~Maza, D.~R. Parisi, L.~A. Pugnaloni,
  I.~Zuriguel, Experimental proof of faster-is-slower in systems of frictional
  particles flowing through constrictions, Phys. Rev. E 92~(6) (2015) 062817
  (Dec. 2015).
\newblock \href {https://doi.org/10.1103/PhysRevE.92.062817}
  {\path{doi:10.1103/PhysRevE.92.062817}}.

\bibitem{nicolas:2018counterintuitive}
A.~Nicolas, S.~Ib\'a\~nez, M.~N. Kuperman, S.~Bouzat, A counterintuitive way to
  speed up pedestrian and granular bottleneck flows prone to clogging: Can
  `more' escape faster?, J. Stat. Mech. 2018~(8) (2018) 083403 (Aug. 2018).
\newblock \href {https://doi.org/10.1088/1742-5468/aad6c0}
  {\path{doi:10.1088/1742-5468/aad6c0}}.

\bibitem{bain:2019dynamic}
N.~Bain, D.~Bartolo, Dynamic response and hydrodynamics of polarized crowds,
  Science 363~(6422) (2019) 46--49 (Jan. 2019).
\newblock \href {https://doi.org/10.1126/science.aat9891}
  {\path{doi:10.1126/science.aat9891}}.

\bibitem{togawa:1955study}
K.~Togawa, Study on Fire Escapes Basing on the Observation of Multitude
  Currents, {Building Research Institute, Ministry of Construction}, 1955
  (1955).

\bibitem{henderson:1971statistics}
L.~F. Henderson, The {{Statistics}} of {{Crowd Fluids}}, Nature 229 (1971)
  381--383 (Feb. 1971).
\newblock \href {https://doi.org/10.1038/229381a0}
  {\path{doi:10.1038/229381a0}}.

\bibitem{helbing:1995social}
D.~Helbing, P.~Moln\'ar, Social force model for pedestrian dynamics, Phys. Rev.
  E 51~(5) (1995) 4282--4286 (May 1995).
\newblock \href {https://doi.org/10.1103/PhysRevE.51.4282}
  {\path{doi:10.1103/PhysRevE.51.4282}}.

\bibitem{vermuyten:2016review}
H.~Vermuyten, J.~Beli\"en, L.~De~Boeck, G.~Reniers, T.~Wauters, A review of
  optimisation models for pedestrian evacuation and design problems, Safety
  Science 87 (2016) 167--178 (Aug. 2016).
\newblock \href {https://doi.org/10.1016/j.ssci.2016.04.001}
  {\path{doi:10.1016/j.ssci.2016.04.001}}.

\bibitem{low:2000statistical}
D.~J. Low, Statistical physics: {{Following}} the crowd, Nature 407~(6803)
  (2000) 465 (Sep. 2000).
\newblock \href {https://doi.org/10.1038/35035192}
  {\path{doi:10.1038/35035192}}.

\bibitem{nicolas:2019mechanical}
A.~Nicolas, M.~Kuperman, S.~Iba\~nez, S.~Bouzat, C.~{Appert-Rolland},
  Mechanical response of dense pedestrian crowds to the crossing of intruders,
  Sci. Rep. 9~(1) (2019) 105 (Jan. 2019).
\newblock \href {https://doi.org/10.1038/s41598-018-36711-7}
  {\path{doi:10.1038/s41598-018-36711-7}}.

\bibitem{burstedde:2001simulation}
C.~Burstedde, K.~Klauck, A.~Schadschneider, J.~Zittartz, Simulation of
  pedestrian dynamics using a two-dimensional cellular automaton, Physica A
  295~(3\textendash{}4) (2001) 507--525 (Jun. 2001).
\newblock \href {https://doi.org/10.1016/S0378-4371(01)00141-8}
  {\path{doi:10.1016/S0378-4371(01)00141-8}}.

\bibitem{weng:2006cellular}
W.~G. Weng, T.~Chen, H.~Y. Yuan, W.~C. Fan, Cellular automaton simulation of
  pedestrian counter flow with different walk velocities, Phys. Rev. E 74~(3)
  (2006) 036102 (Sep. 2006).
\newblock \href {https://doi.org/10.1103/PhysRevE.74.036102}
  {\path{doi:10.1103/PhysRevE.74.036102}}.

\bibitem{guo:2012heterogeneous}
X.~Guo, J.~Chen, Y.~Zheng, J.~Wei, A heterogeneous lattice gas model for
  simulating pedestrian evacuation, Physica A 391~(3) (2012) 582--592 (Feb.
  2012).
\newblock \href {https://doi.org/10.1016/j.physa.2011.07.055}
  {\path{doi:10.1016/j.physa.2011.07.055}}.

\bibitem{patterson:2017clogging}
G.~A. Patterson, P.~I. Fierens, F.~Sangiuliano~Jimka, P.~G. K\"onig,
  A.~Garcimart\'in, I.~Zuriguel, L.~A. Pugnaloni, D.~R. Parisi, Clogging
  {{Transition}} of {{Vibration}}-{{Driven Vehicles Passing}} through
  {{Constrictions}}, Phys. Rev. Lett. 119~(24) (2017) 248301 (Dec. 2017).
\newblock \href {https://doi.org/10.1103/PhysRevLett.119.248301}
  {\path{doi:10.1103/PhysRevLett.119.248301}}.

\bibitem{aguilar:2018collective}
J.~Aguilar, D.~Monaenkova, V.~Linevich, W.~Savoie, B.~Dutta, H.-S. Kuan, M.~D.
  Betterton, M.~a.~D. Goodisman, D.~I. Goldman, Collective clog control:
  {{Optimizing}} traffic flow in confined biological and robophysical
  excavation, Science 361~(6403) (2018) 672--677 (Aug. 2018).
\newblock \href {https://doi.org/10.1126/science.aan3891}
  {\path{doi:10.1126/science.aan3891}}.

\bibitem{delarue:2016selfdriven}
M.~Delarue, J.~Hartung, C.~Schreck, P.~Gniewek, L.~Hu, S.~Herminghaus,
  O.~Hallatschek, Self-driven jamming in growing microbial populations, Nat.
  Phys. 12~(8) (2016) 762--766 (Aug. 2016).
\newblock \href {https://doi.org/10.1038/nphys3741}
  {\path{doi:10.1038/nphys3741}}.

\bibitem{garcimartin:2015flow}
A.~Garcimart\'in, J.~M. Pastor, L.~M. Ferrer, J.~J. Ramos,
  C.~{Mart\'in-G\'omez}, I.~Zuriguel, Flow and clogging of a sheep herd passing
  through a bottleneck, Phys. Rev. E 91~(2) (2015) 022808 (Feb. 2015).
\newblock \href {https://doi.org/10.1103/PhysRevE.91.022808}
  {\path{doi:10.1103/PhysRevE.91.022808}}.

\bibitem{moussaid:2011how}
M.~Moussa\"id, D.~Helbing, G.~Theraulaz, How simple rules determine pedestrian
  behavior and crowd disasters, PNAS 108~(17) (2011) 6884--6888 (Apr. 2011).
\newblock \href {https://doi.org/10.1073/pnas.1016507108}
  {\path{doi:10.1073/pnas.1016507108}}.

\bibitem{corbetta:2017fluctuations}
A.~Corbetta, C.-m. Lee, R.~Benzi, A.~Muntean, F.~Toschi, Fluctuations around
  mean walking behaviors in diluted pedestrian flows, Phys. Rev. E 95~(3)
  (2017) 032316 (Mar. 2017).
\newblock \href {https://doi.org/10.1103/PhysRevE.95.032316}
  {\path{doi:10.1103/PhysRevE.95.032316}}.

\bibitem{helbing:2011recognition}
D.~Helbing, G.~Tr\"oster, M.~Wirz, D.~Roggen, Recognition of crowd behavior
  from mobile sensors with pattern analysis and graph clustering methods, Netw.
  Heterog. Media 6~(3) (2011) 521--544 (Aug. 2011).
\newblock \href {https://doi.org/10.3934/nhm.2011.6.521}
  {\path{doi:10.3934/nhm.2011.6.521}}.

\bibitem{zanlungo:2017intrinsic}
F.~Zanlungo, Z.~Yucel, D.~Brscic, T.~Kanda, N.~Hagita, Intrinsic group
  behaviour: Dependence of pedestrian dyad dynamics on principal social and
  personal features, PLOS ONE 12~(11) (2017) e0187253 (Nov. 2017).
\newblock \href {http://arxiv.org/abs/1703.02672} {\path{arXiv:1703.02672}},
  \href {https://doi.org/10.1371/journal.pone.0187253}
  {\path{doi:10.1371/journal.pone.0187253}}.

\bibitem{wang:2018study}
C.~Wang, W.~Weng, Study on the collision dynamics and the transmission pattern
  between pedestrians along the queue, J. Stat. Mech. 2018~(7) (2018) 073406
  (Jul. 2018).
\newblock \href {https://doi.org/10.1088/1742-5468/aace27}
  {\path{doi:10.1088/1742-5468/aace27}}.

\bibitem{shi:2018examining}
X.~Shi, Z.~Ye, N.~Shiwakoti, D.~Tang, J.~Lin, Examining effect of architectural
  adjustment on pedestrian crowd flow at bottleneck, ArXiv180807439 Phys. (Aug.
  2018).
\newblock \href {http://arxiv.org/abs/1808.07439} {\path{arXiv:1808.07439}}.

\bibitem{kulkarni:2019sparse}
A.~Kulkarni, S.~P. Thampi, M.~V. Panchagnula, Sparse {{Game Changers Restore
  Collective Motion}} in {{Panicked Human Crowds}}, Phys. Rev. Lett. 122~(4)
  (2019) 048002 (Jan. 2019).
\newblock \href {https://doi.org/10.1103/PhysRevLett.122.048002}
  {\path{doi:10.1103/PhysRevLett.122.048002}}.

\bibitem{dongmei:2017dynamics}
S.~Dongmei, Z.~Wenyao, W.~Binghong, Dynamics of {{Panic Pedestrians}} in
  {{Evacuation}}, ArXiv170101236 Nlin Physicsphysics (Jan. 2017).
\newblock \href {http://arxiv.org/abs/1701.01236} {\path{arXiv:1701.01236}}.

\bibitem{hao:2011pedestrian}
Q.-Y. Hao, R.~Jiang, M.-B. Hu, B.~Jia, Q.-S. Wu, Pedestrian flow dynamics in a
  lattice gas model coupled with an evolutionary game, Phys. Rev. E 84~(3)
  (2011) 036107 (Sep. 2011).
\newblock \href {https://doi.org/10.1103/PhysRevE.84.036107}
  {\path{doi:10.1103/PhysRevE.84.036107}}.

\bibitem{zheng:2011conflict}
X.~Zheng, Y.~Cheng, Conflict game in evacuation process: {{A}} study combining
  {{Cellular Automata}} model, Physica A: Statistical Mechanics and its
  Applications 390~(6) (2011) 1042--1050 (Mar. 2011).
\newblock \href {https://doi.org/10.1016/j.physa.2010.12.007}
  {\path{doi:10.1016/j.physa.2010.12.007}}.

\bibitem{zheng:2011modeling}
X.~Zheng, Y.~Cheng, Modeling cooperative and competitive behaviors in emergency
  evacuation: {{A}} game-theoretical approach, Computers \& Mathematics with
  Applications 62~(12) (2011) 4627--4634 (Dec. 2011).
\newblock \href {https://doi.org/10.1016/j.camwa.2011.10.048}
  {\path{doi:10.1016/j.camwa.2011.10.048}}.

\bibitem{shi:2013evacuation}
D.-M. Shi, B.-H. Wang, Evacuation of pedestrians from a single room by using
  snowdrift game theories, Phys. Rev. E 87~(2) (2013) 022802 (Feb. 2013).
\newblock \href {https://doi.org/10.1103/PhysRevE.87.022802}
  {\path{doi:10.1103/PhysRevE.87.022802}}.

\bibitem{guan:2016cellular}
J.~Guan, K.~Wang, F.~Chen, A cellular automaton model for evacuation flow using
  game theory, Physica A: Statistical Mechanics and its Applications 461 (2016)
  655--661 (Nov. 2016).
\newblock \href {https://doi.org/10.1016/j.physa.2016.05.062}
  {\path{doi:10.1016/j.physa.2016.05.062}}.

\bibitem{vonschantz:2015spatial}
A.~{von Schantz}, H.~Ehtamo, Spatial game in cellular automaton evacuation
  model, Phys. Rev. E 92~(5) (2015) 052805 (Nov. 2015).
\newblock \href {https://doi.org/10.1103/PhysRevE.92.052805}
  {\path{doi:10.1103/PhysRevE.92.052805}}.

\bibitem{nicolas:2018trap}
A.~Nicolas, A.~Garcimart\'in, I.~Zuriguel, Trap {{Model}} for {{Clogging}} and
  {{Unclogging}} in {{Granular Hopper Flows}}, Phys. Rev. Lett. 120~(19) (2018)
  198002 (May 2018).
\newblock \href {https://doi.org/10.1103/PhysRevLett.120.198002}
  {\path{doi:10.1103/PhysRevLett.120.198002}}.

\bibitem{cavagna:2018physics}
A.~Cavagna, I.~Giardina, T.~S. Grigera, The physics of flocking:
  {{Correlation}} as a compass from experiments to theory, Physics Reports 728
  (2018) 1--62 (Jan. 2018).
\newblock \href {https://doi.org/10.1016/j.physrep.2017.11.003}
  {\path{doi:10.1016/j.physrep.2017.11.003}}.

\bibitem{castellano:2009statistical}
C.~Castellano, S.~Fortunato, V.~Loreto, Statistical physics of social dynamics,
  Rev. Mod. Phys. 81~(2) (2009) 591--646 (May 2009).
\newblock \href {https://doi.org/10.1103/RevModPhys.81.591}
  {\path{doi:10.1103/RevModPhys.81.591}}.

\bibitem{ball:2012why}
P.~Ball, Why {{Society}} Is a {{Complex Matter}}: {{Meeting Twenty}}-First
  {{Century Challenges}} with a {{New Kind}} of {{Science}}, {Springer}, Berlin
  Heidelberg, 2012 (2012).

\bibitem{nowak:2005emergence}
A.~Nowak, R.~R. Vallacher, M.~Zochowski, The emergence of personality:
  {{Dynamic}} foundations of individual variation, Developmental Review
  25~(3\textendash{}4) (2005) 351--385 (Sep. 2005).
\newblock \href {https://doi.org/10.1016/j.dr.2005.10.004}
  {\path{doi:10.1016/j.dr.2005.10.004}}.

\bibitem{pan:2014spatial}
Q.~Pan, L.~Wang, R.~Shi, H.~Wang, H.~Mingfeng, Spatial modes of cooperation
  based on bounded rationality, Physica A 415 (2014) 421--427 (Dec. 2014).
\newblock \href {https://doi.org/10.1016/j.physa.2014.07.058}
  {\path{doi:10.1016/j.physa.2014.07.058}}.

\bibitem{simon:1983reason}
H.~A. Simon, Reason in Human Affairs, {Stanford Univ. Press}, Stanford, Calif,
  1983 (1983).

\bibitem{heliovaara:2013patient}
S.~Heli\"ovaara, H.~Ehtamo, D.~Helbing, T.~Korhonen, Patient and impatient
  pedestrians in a spatial game for egress congestion, Phys. Rev. E 87~(1)
  (2013) 012802 (Jan. 2013).
\newblock \href {https://doi.org/10.1103/PhysRevE.87.012802}
  {\path{doi:10.1103/PhysRevE.87.012802}}.

\bibitem{taylor:1978evolutionary}
P.~D. Taylor, L.~B. Jonker, Evolutionary stable strategies and game dynamics,
  Mathematical Biosciences 40~(1) (1978) 145--156 (Jul. 1978).
\newblock \href {https://doi.org/10.1016/0025-5564(78)90077-9}
  {\path{doi:10.1016/0025-5564(78)90077-9}}.

\bibitem{lasry:2007mean}
J.-M. Lasry, P.-L. Lions, Mean field games, Jpn. J. Math. 2~(1) (2007) 229--260
  (Mar. 2007).
\newblock \href {https://doi.org/10.1007/s11537-007-0657-8}
  {\path{doi:10.1007/s11537-007-0657-8}}.

\bibitem{kirchner:2003friction}
A.~Kirchner, K.~Nishinari, A.~Schadschneider, Friction effects and clogging in
  a cellular automaton model for pedestrian dynamics, Phys. Rev. E 67~(5)
  (2003) 056122 (May 2003).
\newblock \href {https://doi.org/10.1103/PhysRevE.67.056122}
  {\path{doi:10.1103/PhysRevE.67.056122}}.

\bibitem{lee:2001effects}
K.~Lee, P.~M. Hui, B.-H. Wang, N.~F. Johnson, Effects of {{Announcing Global
  Information}} in a {{Two}}-{{Route Traffic Flow Model}}, J. Phys. Soc. Jpn.
  70~(12) (2001) 3507--3510 (Dec. 2001).
\newblock \href {https://doi.org/10.1143/JPSJ.70.3507}
  {\path{doi:10.1143/JPSJ.70.3507}}.

\bibitem{mao:2016experimental}
A.~Mao, W.~Mason, S.~Suri, D.~J. Watts, An {{Experimental Study}} of {{Team
  Size}} and {{Performance}} on a {{Complex Task}}, PLOS ONE 11~(4) (2016)
  e0153048 (Apr. 2016).
\newblock \href {https://doi.org/10.1371/journal.pone.0153048}
  {\path{doi:10.1371/journal.pone.0153048}}.

\bibitem{awad:2018moral}
E.~Awad, S.~Dsouza, R.~Kim, J.~Schulz, J.~Henrich, A.~Shariff, J.-F. Bonnefon,
  I.~Rahwan, The {{Moral Machine}} experiment, Nature 563~(7729) (2018) 59
  (Nov. 2018).
\newblock \href {https://doi.org/10.1038/s41586-018-0637-6}
  {\path{doi:10.1038/s41586-018-0637-6}}.

\end{thebibliography}

\end{document}